\documentclass[twocolumn,showpacs,superscriptaddress,prl]{revtex4}
\usepackage{amssymb}
\usepackage{graphicx} 

\begin{document}

\title{Double and negative reflection of cold atoms in non-Abelian gauge
potentials}

\date{\today{}}

\author{Gediminas Juzeli\={u}nas}
\affiliation{Institute of Theoretical Physics and Astronomy of Vilnius
University, A. Go\v{s}tauto 12, Vilnius 01108, Lithuania}
\affiliation{Department of Physics and Technology,
Vilnius Pedagogical University, Studentu 39, Vilnius 08106, Lithuania}

\author{Julius Ruseckas}
\affiliation{Institute of Theoretical Physics and Astronomy of Vilnius
University, A. Go\v{s}tauto 12, Vilnius 01108, Lithuania}

\author{Andreas Jacob}
\affiliation{Institut f\"{u}r Theoretische Physik, Leibniz Universit\"{a}t,
Hannover D30167, Germany}

\author{Luis Santos}
\affiliation{Institut f\"{u}r Theoretische Physik, Leibniz Universit\"{a}t,
Hannover D30167, Germany}

\author{Patrik \"{O}hberg}
\affiliation{SUPA, School of Engineering and Physical Sciences, Heriot-Watt University, Edinburgh
EH14 4AS, United Kingdom}

\begin{abstract}
Atom reflection is studied in the presence of a non-Abelian vector 
potential proportional to a spin-$1/2$ operator. The potential is produced by a relatively 
simple laser configuration for atoms with a tripod level scheme.  We show 
that the atomic motion is described by two different dispersion branches with 
positive or negative chirality. As a consequence, atom reflection 
shows unusual features, since an incident wave may split into two 
reflected ones at a barrier, an ordinary specular reflection, and an 
additional non-specular one. Remarkably, the latter wave can exhibit negative 
reflection and may become evanescent if the angle of incidence exceeds a 
critical value. These reflection properties are crucial for future designs 
in non-Abelian atom optics.
\end{abstract}


\maketitle

\paragraph{Introduction}

Atomic mirrors, created by optical \cite{OPT} or magnetic \cite{MAGN} 
potential barriers, play a
crucial role in atom optics enabling to manipulate matter waves.
Wave reflection at a mirror is typically specular, where 
the reflection angle equals the incidence one. However, richer reflection
scenarios are also possible.  For optical waves a double reflection appears in
optically active media, such as in chiral liquids characterized by different
refractive indices for left and right polarized light \cite{barron04}.  This
manifests itself as a tiny splitting of the reflected wave into two parts
\cite{ghosh06}. An additional striking example is Andreev reflection
\cite{andreev64,beenakker06} in which an electron incident at the interface
between a normal metal and a superconductor is reflected to a positively
charged hole propagating backwards, where the missing charge of $2e$ enters the
superconductor as a Cooper pair.

Artificial electromagnetism for cold neutral atoms is attracting a
growing attention. A proper manipulation of atoms in optical lattices may allow
for the observation of Hofstadter butterfly energy spectra \cite{Jaksch} and
paradoxical geometries \cite{Mueller04}, as well as for the generation of
non-Abelian gauge potentials \cite{osterloh05}. Alternatively, Abelian and
non-Abelian gauge potentials may be induced by means of laser fields acting on
atoms in, respectively, lambda \cite{Lambda} and tripod
\cite{ruseckas05,stanescu07,clark07}  schemes of electronic levels.
Non-Abelian gauge potentials lead to a number of distinctive properties, e.g.
modification of the metal-insulator transition \cite{Satija06} or non-Abelian
Aharanov-Bohm interferometric effects \cite{osterloh05,Jacob07}. Recently, it
was shown \cite{prar08} that a tripod-scheme provides quasi-relativistic
physics for cold atoms, similar to that for electrons in graphene
\cite{Graphene}.

In this Letter we analyze atom reflection in the presence of a non-Abelian
vector potential proportional to a spin-$1/2$ operator produced by a
relatively simple laser arrangement for tripod-scheme atoms. 
We show that the
appearance of two different dispersion branches with positive or negative
chirality leads to a double reflection at the mirror, an ordinary specular
reflection, and an additional non-specular one. Remarkably, the latter can
exhibit a negative reflection, resembling the Andreev reflection
\cite{andreev64,beenakker06}. The negatively reflected wave becomes evanescent
if the angle of incidence exceeds a critical value.  These reflection
properties could become crucial for the design of future non-Abelian atom
optics devices, as e.g.\ non-Abelian atom interferometers \cite{osterloh05,Jacob07}.

\paragraph{Adiabatic motion of tripod atoms}

\begin{figure}
\centering
\includegraphics[width=0.45\textwidth]{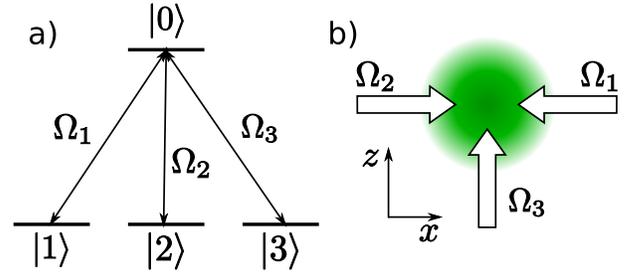}
\caption{(Color online) Three light fields acting on atom in a tripod
configuration of energy levels involved.}
\label{fig:tripod}
\end{figure}

In the following we consider an atom with a tripod electronic level scheme
$\{|0\rangle, |1\rangle, |2\rangle, |3\rangle\}$ (see Fig.~\ref{fig:tripod}a)
under the influence of three stationary laser beams
\cite{ruseckas05,prar08,unanyan98,unanyan99}.  The $j$-th laser induces a
resonant atomic transition (with a Rabi frequency $\Omega_j$) between
$|j\rangle$ and $|0\rangle$. These can be for instance the transition
$2^3S_1\leftrightarrow2^3P_0$ in $^4\mathrm{He}^*$ or the transition
$5S_{1/2}(F=1)\leftrightarrow5P_{3/2}(F=0)$ in $^{87}\mathrm{Rb}$.

The electronic Hamiltonian of the tripod system reads in the interaction
representation \cite{ruseckas05}
\begin{equation}
\hat{H}_0=-\hbar|0\rangle\Bigl(\Omega_1\langle1|+\Omega_2\langle2|
+\Omega_3\langle3|\Bigr)+\mathrm{H.c.}\,.
\label{eq:H-0}
\end{equation}
Thus $|0\rangle$ is coupled only to the bright state
$|B\rangle=(\Omega_1^*|1\rangle+\Omega_2^*|2\rangle+\Omega_3^*|3\rangle)/\Omega$,
where $\Omega=(|\Omega_1|^2+|\Omega_2|^2+|\Omega_3|^2)^{1/2}$ is the total Rabi
frequency. The two states $|B\rangle$ and $|0\rangle$ split into a dressed
doublet $|\pm\rangle=(|B\rangle\pm|0\rangle)/\sqrt{2}$ with energies
$\pm\hbar\Omega$. The remaining two eigenstates $|D_j\rangle$ ($j=1\,,2$),
known as dark states,  are orthogonal to $|B\rangle$ and hence are decoupled
from the light fields, $\hat{H}_0|D_j\rangle=0$. We assume that the  light
fields are sufficiently strong, so that $\Omega$ is large compared to the
two-photon detuning due to the laser mismatch and/or Doppler shift. The dark
states are thus well separated in energies from the doublet $|\pm\rangle$, and
the internal atomic state evolves within the dark state manifold. The full
atomic state-vector $|\Phi\rangle$ can then be expanded in terms of the dark
states, $|\Phi\rangle=\sum_{j=1}^2\Psi_j(\mathbf{r})|D_j(\mathbf{r})\rangle$,
where $\Psi_j(\mathbf{r})$ is a wave-function for the center of mass motion of
an atom in the $j$-th dark state. The two-component spinor-like wavefunction
$\Psi=\{\Psi_1,\Psi_2\}^T$ obeys the Schr\"{o}dinger equation
$i\hbar\partial\Psi/\partial t=H\Psi$, with the center of mass Hamiltonian
\cite{ruseckas05}
\begin{equation}
H=\frac{1}{2M}(-i\hbar\nabla-\mathbf{A})^2+V+\Phi\,,
\end{equation}
where $M$ is the atomic mass. The gauge potentials $\mathbf{A}$ and $\Phi$
emerge due to the spatial dependence of the dark states. The $2\times2$ matrix
$\mathbf{A}$ with the elements $\mathbf{A}_{nm}=i\hbar\langle
D_n(\mathbf{r})|\nabla D_m(\mathbf{r})\rangle$ represents the effective vector
potential known as the Mead-Berry connection
\cite{unanyan99,ruseckas05,prar08,berry84,wilczek84,mead91}. The $2\times2$
matrix $\Phi$ with elements $\Phi_{nm}=(\hbar^2/2M)\langle
D_n(\mathbf{r})|\nabla B(\mathbf{r})\rangle\langle B(\mathbf{r})|\nabla
D_m(\mathbf{r})\rangle$ acts as an effective scalar potential. Finally, the
$2\times2$ matrix $V$ with elements $V_{nm}=\langle
D_n(\mathbf{r})|\hat{V}|D_m(\mathbf{r})\rangle$ is an external potential for
the dark-state atoms. Here $\hat{V}=\sum_{j=1}^3V_j(\mathbf{r})|j\rangle\langle
j|$, with $V_j(\mathbf{r})$ being the trapping potential for an atom in the
$j$-the internal state. Note that the potential $V_j$ can also accommodate a
detuning of the $j$-th laser from the resonance of the $j\rightarrow0$
transition.

\paragraph{Laser arrangement}

Although elaborate laser configurations may allow for a wealth of possible
gauge potentials in the tripod scheme \cite{ruseckas05}, here we concentrate on
a relatively simple laser setup providing non Abelian potentials. The
first two laser beams are assumed to counterpropagate with the same intensity  along the
$x$-axis, $\Omega_1=\Omega\sin\theta\mathrm{e}^{-i\kappa_0x}/\sqrt{2}$ and
$\Omega_2=\Omega\sin\theta\mathrm{e}^{i\kappa_0x}/\sqrt{2}$, and the third one
propagating in the $z$-direction \cite{prar08},
$\Omega_3=\Omega\cos\theta\mathrm{e}^{i\kappa_0z}$. Here $\kappa_0$ is the
wave-number, and the mixing angle $\theta$ characterizes the relative intensity
of the third laser. A set of two dark states is then given by 
\begin{eqnarray}
|D_1\rangle & = &
2^{-1/2}\left(|\tilde{1}\rangle-|\tilde{2}\rangle\right)
e^{-i\kappa^{\prime}z}\,,
\label{eq:D1-plane-wave}
\\ |D_2\rangle & = &
\left[2^{-1/2}\cos\theta\left(|\tilde{1}\rangle+|\tilde{2}\rangle\right)
-\sin\theta|3\rangle\right]e^{-i\kappa^{\prime}z}\,,
\label{eq:D2-plane-wave}
\end{eqnarray}
with $\kappa^{\prime}=\kappa_0(1-\cos\theta)$, where the modified atomic
state-vectors $|\tilde{1}\rangle=|1\rangle\exp[i\kappa_0(x+z)]$ and
$|\tilde{2}\rangle=|2\rangle\exp[-i\kappa_0(x-z)]$ accommodate the phases of
the laser fields. An additional phase factor $\exp(i\kappa^{\prime}z)$
introduces a shift in the origin of the momentum
$\mathbf{k}\rightarrow\mathbf{k}+\kappa^{\prime}\mathbf{e}_z$. By imposing
$\cos\theta=\sqrt{2}-1$, the vector potential becomes
$\mathbf{A}=-\hbar\kappa\mathbf{\sigma}_{\bot}$, where
$\mathbf{\sigma}_{\bot}=\mathbf{e}_x\sigma_x+\mathbf{e}_z\sigma_z$ is the
operator of spin $1/2$ in the $xz$ plane, and
$\kappa=\kappa_0\cos\theta\approx0.414\kappa_0$. The Cartesian components
$A_x$ and $A_z$ are proportional to the Pauli matrices $\sigma_x$ and
$\sigma_z$ which do not commute, so the vector potential $\mathbf{A}$ is
non-Abelian. Note that similar non-Abelian gauge potentials can also be induced
by means of standing waves in a tripod setup \cite{stanescu07,clark07} or by
state selective tunneling in optical lattices \cite{osterloh05}.

Furthermore we take the trapping potentials $V_1=V_2$ and
$V_3-V_1=\hbar\kappa_0^2\sin^2\theta/2M$. This can be achieved by detuning
properly the third laser from the two-photon resonance. Hence the overall
trapping potential $V+\Phi$ becomes proportional to the unit matrix, both dark
states being affected by the same potential $V_1\equiv V_1(\mathbf{r})$, giving
\begin{equation}
H=\frac{1}{2M}(-i\hbar\nabla+\hbar\kappa\mathbf{\sigma}_{\bot})^2+
V_1(\mathbf{r})\,.
\label{eq:H-cm}
\end{equation}

\paragraph{Dispersion law}

We shall consider a two dimensional case where the atomic motion is confined to
the $xz$ plane. If the trapping potential $V_1$ is constant, the eigenfunctions
of the Hamiltonian (\ref{eq:H-cm}) are the plane waves 
\begin{equation}
\Psi_{\mathbf{k}}^{\pm}\left(\mathbf{r}\right)=g_{\mathbf{k}}^{\pm}e^{i\mathbf{
k}\cdot\mathbf{r}}\,,\qquad g_{\mathbf{k}}^{\pm}=\frac{1}{2}\left(
\begin{array}{c}
1\mp ie^{i\varphi_{\mathbf{k}}}\\
-i\pm e^{i\varphi_{\mathbf{k}}}
\end{array}\right)\,,
\label{eq:Psi-k-pm}
\end{equation}
where $\varphi_{\mathbf{k}}$ is the angle between the atomic wave-vector
$\mathbf{k}$ and the $x$-axis. The two-component spinors $g_{\mathbf{k}}^{\pm}$
are eigenfunctions of the chirality operator
$\sigma_{\mathbf{k}}=\sigma\cdot\mathbf{k}/k$ representing a spin along the
atomic motion, $\sigma_{\mathbf{k}}g_{\mathbf{k}}^{\pm}=\pm
g_{\mathbf{k}}^{\pm}$. It should be emphasized that the chirality is here
associated with the subspace of two dark states rather than with the spin in
the usual sense.

\begin{figure}
\centering
\includegraphics[width=0.45\textwidth]{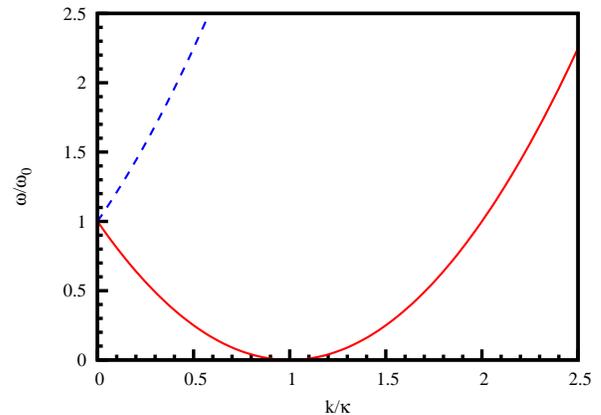}
\caption{(Color online) Upper (blue dashed) and lower (red solid)
dispersion branch for tripod atoms in light fields.}
\label{fig:dispersion}
\end{figure}

The corresponding eigenenergies of the Hamiltonian (\ref{eq:H-cm}) are
isotropic, $\hbar\omega_{\mathbf{k}}^{\pm}\equiv\hbar\omega_k^{\pm}$, with
\begin{equation}
\hbar\omega_k^{\pm}=\frac{\hbar^2}{2M}(k\pm\kappa)^2+\hbar\omega_0+V_1,
\label{eq:omega-k-pm}
\end{equation}
where $\omega_0=\hbar\kappa^2/2M$ is the recoil frequency.  The relative
dispersion $\omega_k^{\pm}/\omega_0$ is plotted in Fig.~\ref{fig:dispersion}
for $V_1=-\hbar\omega_0$. The upper (lower) dispersion branch is characterized by a
positive (negative) chirality. For small wave-numbers $k\ll\kappa$ the
dispersion is linear, $\omega_k^{\pm}\sim\pm k$, so the atoms behave like
ultra-relativistic Dirac fermions \cite{prar08} similar to electrons in
graphene \cite{Graphene}. For larger wave-numbers each $k=k_1<2\kappa$ has a
counterpart, $k_2=2\kappa-k$, characterized by the same eigen-frequency
$\hbar\omega_{k_2}^{-}=\hbar\omega_{k_1}^{-}$ and opposite slope in the lower
dispersion branch \cite{pendry04}. As shown below, this leads to unusual
reflection properties.

\paragraph{Double and negative reflection of atoms}

To prepare an incident atom in the lower dispersion branch we suggest the following procedure.
Initially the first two lasers are on, the third laser is off, and the atom is in the
internal state $|3\rangle$ coinciding (up to a phase) with the second dark state, $|D_2\rangle=|3\rangle\exp(-i\kappa^{\prime}z)$.
The atomic center of mass motion is initially characterized by a wave-vector
$\mathbf{k}_{\mathrm{in}}=k_{\mathrm{in}}\mathbf{e}_z$ \cite{Deng99}, giving the full state
vector $|\Phi\rangle=|D_2\rangle\exp(ikz)$, where
$k=k_{\mathrm{in}}+\kappa^{\prime}$.
Subsequently the laser $3$ is switched on slowly, so that the atom remains in the dark 
state $|D_2\rangle$. 
%which however becomes different from the physical state $|3\rangle$. 
Yet the duration of the switching-on should be short enough
to avoid any substantial atomic motion at this stage.
Since $\mathbf{k}=k \mathbf{e}_z$, the spinor
$g_{\bf k}^{-}$ represents the dark state $|D_2\rangle$. Hence one arrives at the atomic state-vector $|\Phi\rangle$ corresponding to
the negative-chirality solution $\Psi_{\mathbf{k}}^{-}$, as required. The atoms prepared in this way
will propagate along the $z$ axis for $k>\kappa$ or opposite to it for
$k<\kappa$.  

\begin{figure}
\centering
\includegraphics[width=0.2\textwidth]{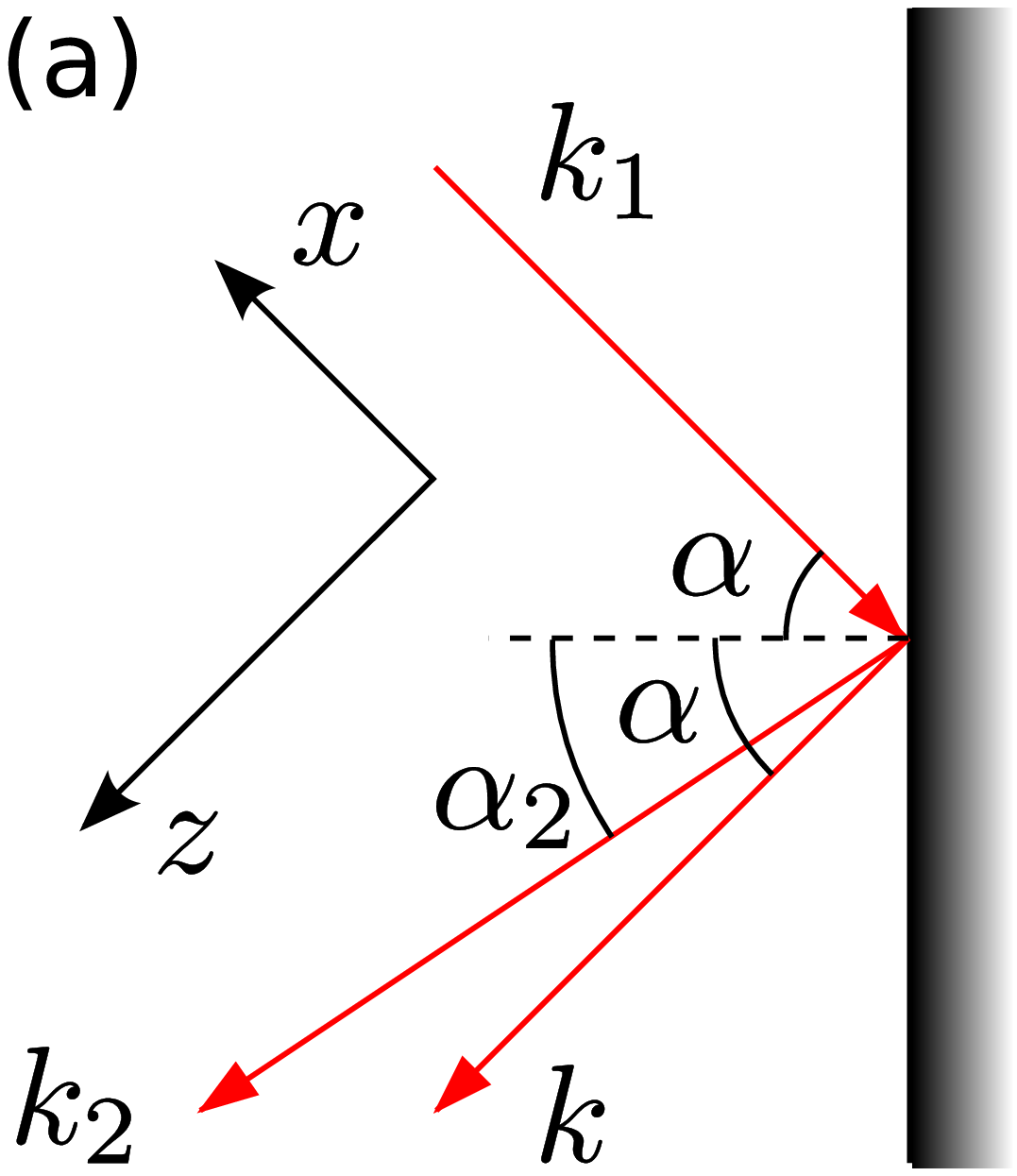}\includegraphics[ width=0.2\textwidth]{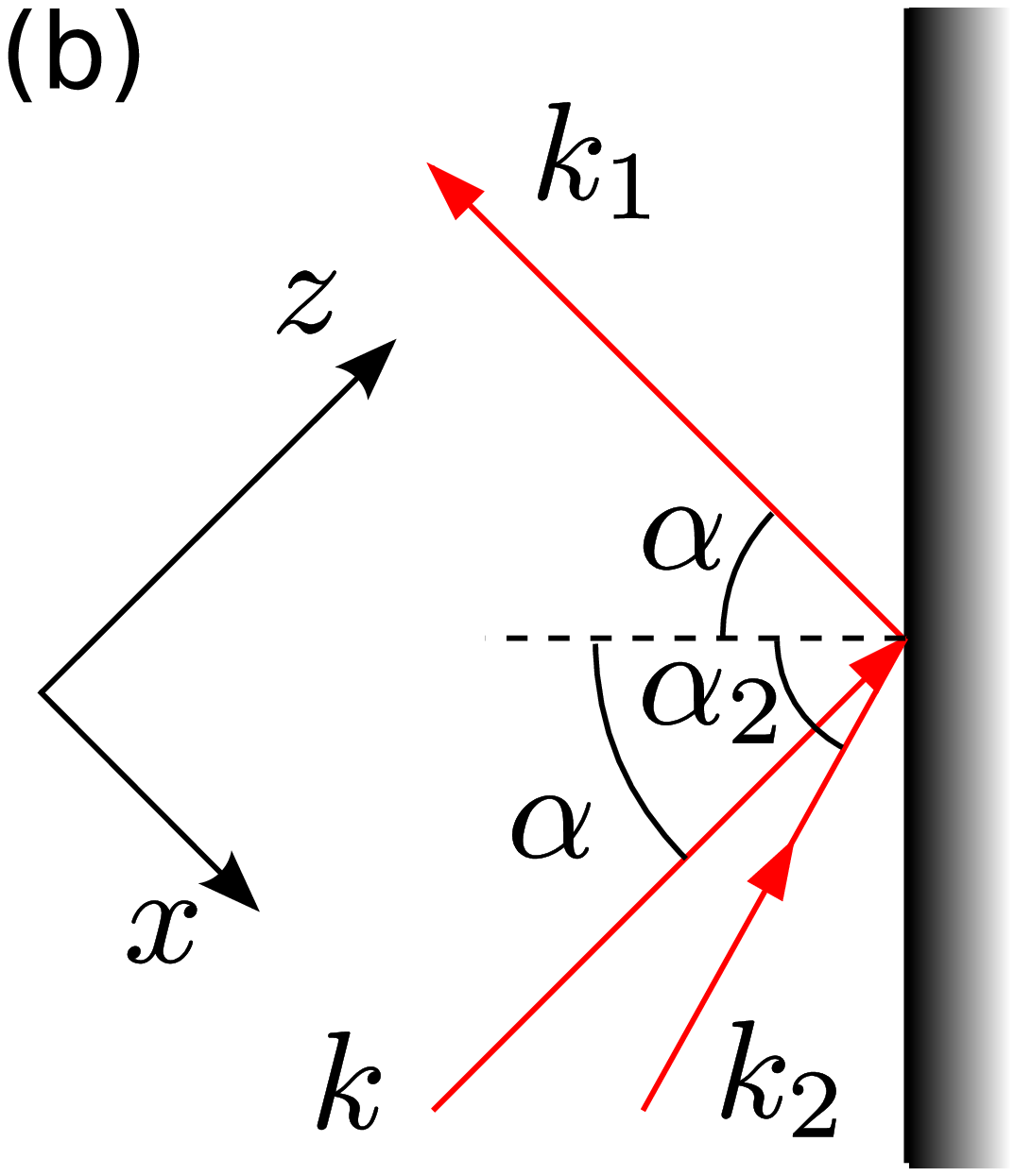}
\caption{(Color online) Reflection of atoms with negative
chirality for $0<k<\kappa$ (a) and $\kappa<k<2\kappa$ (b).}
\label{fig:reflection}
\end{figure}

The atoms are impinging on an infinitely high potential barrier at an angle of
incidence $\alpha$.  We shall take $k<2\kappa$, so that both reflected waves
$\Psi_{\mathbf{k}_1}^{-}$ and $\Psi_{\mathbf{k}_2}^{-}$ remain in the lower
dispersion branch with wave-numbers $k_1=k$ and $k_2=2\kappa-k$.
Fig.~\ref{fig:reflection}b shows the case of an incident wave with
$\kappa<k<2\kappa$. Here the group velocity
$v_k^{-}=\partial\omega_k^{-}/\partial k$ is positive, so the wave-vectors of
the incident and second reflected waves point inwards to the surface, whereas
the wave-vector of the first reflected wave points outwards from the surface.
Since $v_{k}^{-}=v_{k_1}^{-}=-v_{k_2}^{-}$, this ensures the forward
propagation of the incident wave and backward propagation of the reflected
ones. Fig.~\ref{fig:reflection}a illustrates a situation where $0<k<\kappa$.
Here the group velocity $v_k^{-}$ is negative and hence the wave-vectors are
reversed.

\begin{figure}
\centering
\includegraphics[width=0.45\textwidth]{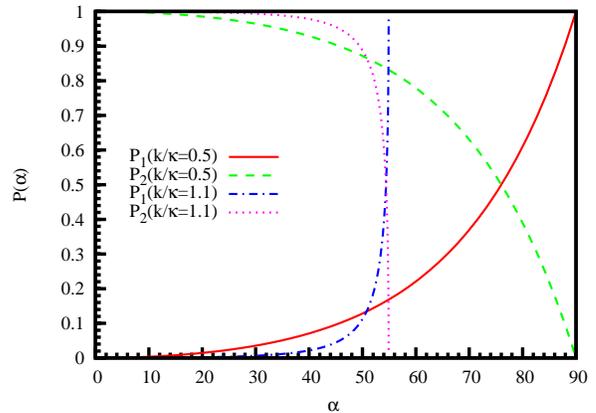}
\caption{(Color online) Reflection probabilities $P_1$ and $P_2$ for
$k/\kappa=0.5$ and $k/\kappa=1.1$.}
\label{fig:prob}
\end{figure}

In front of the barrier the solution to the stationary Schr\"{o}dinger equation
$(H-\hbar\omega_k^{-})\Psi=0$ is a linear superposition of the incident wave
and two reflected waves: 
\begin{equation}
\Psi=\Psi_{\mathbf{k}}^{-}+r_1\Psi_{\mathbf{k}_1}^{-}+
r_2\Psi_{\mathbf{k}_2}^{-}\,.
\label{eq:psi-full}
\end{equation}
The wave-vector is conserved along the reflection plane,
$k_{\Vert}=k_{1\Vert}=k_{2\Vert}$, so the first wave exhibits an ordinary
reflection with the reflection angle equal to the angle of incidence,
$\alpha_1=\alpha$. The second wave is characterized by the opposite group
velocity $v_{k_2}^{-}=-v_{k}^{-}=-v_{k_1}^{-}$, and hence it experiences a
\textit{negative reflection} at an angle 
\begin{equation}
\alpha_2=\arcsin[\sin(\alpha)k/k_2]\,.
\label{eq:alpha-2}
\end{equation}
The reflection coefficients $r_1$ and $r_2$ are determined using
Eqs.~(\ref{eq:Psi-k-pm}) and (\ref{eq:psi-full}) together with the boundary
condition at the potential barrier $\Psi|_{\mathrm{barrier}}=0$, giving
\begin{equation}
r_1=\frac{e^{i\alpha}-e^{i\alpha_2}}{e^{-i\alpha}+e^{i\alpha_2}}\,,\qquad
r_2=-1-r_1\,.
\label{eq:r-1-r-2}
\end{equation}
The corresponding reflection probabilities are 
\begin{equation}
P_1=|r_1|^2,\qquad
P_2=\frac{\cos\alpha_2}{\cos\alpha}|r_2|^2\,,
\label{eq:P-1-P-2}
\end{equation}
with $P_1+P_2=1$, where the weight factor $\cos\alpha_2/\cos\alpha$ appears
when calculating the flow of atoms in and out of the surface for the incident
and reflected waves. The probabilities $P_1$ and $P_2$ plotted in
Figure~\ref{fig:prob} are seen to depend both on the angle of incidence
$\alpha$ and also on the wave-number $k$. For small angles, $\alpha\ll1$, there
is predominantly a negative reflection to the second branch, $|P_1|\ll1$ and
$|P_2|\approx1$. For large angles of incidence ($\alpha\rightarrow\pi/2$) and
$0<k<\kappa$ we have mostly a specular reflection to the first branch,
$|P_2|\ll1$ and $|P_1|\approx1$. 

If $\kappa<k<2\kappa$, the situation is more complex. In this case the second
reflected wave becomes evanescent when the angle of incidence $\alpha$ exceeds
a critical value given by $\sin\alpha_{\mathrm{crit}}=k_2/k$, i.e. for
$k_{\Vert}=k_{2\Vert}>k_2$. Consequently the out-of-plane projection of the
wave-vector $\mathbf{k}_2$ becomes imaginary, $k_{2\bot}=-iq$, with
$q=\sqrt{k_{\Vert}^2-k_2^2}$. In the region where $x<0$ we can once again use
Eq.~(\ref{eq:psi-full}) in which $\Psi_{\mathbf{k}_2}^{-}$ is now an evanescent
wave. The boundary condition at the potential barrier gives the reflection
coefficient 
\begin{equation}
r_1=\frac{e^{i\alpha}\sqrt{k_{\Vert}+q}-i\sqrt{k_{\Vert}-q}}{e^{-i\alpha}\sqrt{
k_{\Vert}+q}+i\sqrt{k_{\Vert}-q}}\,,
\label{eq:r-1-evanescent}
\end{equation}
with $|r_1|=1$. Thus there is a total reflection to the first mode at an angle
$\alpha_1=\alpha$ accompanied by a phase shift, with the second reflected wave
being evanescent. The phenomenon resembles the total internal reflection of
optical waves at an interface with an optically thinner medium. In our
situation, however, the evanescent wave is the reflected wave rather than the
refracted one.

\begin{figure}
\centering
\includegraphics[width=0.23\textwidth]{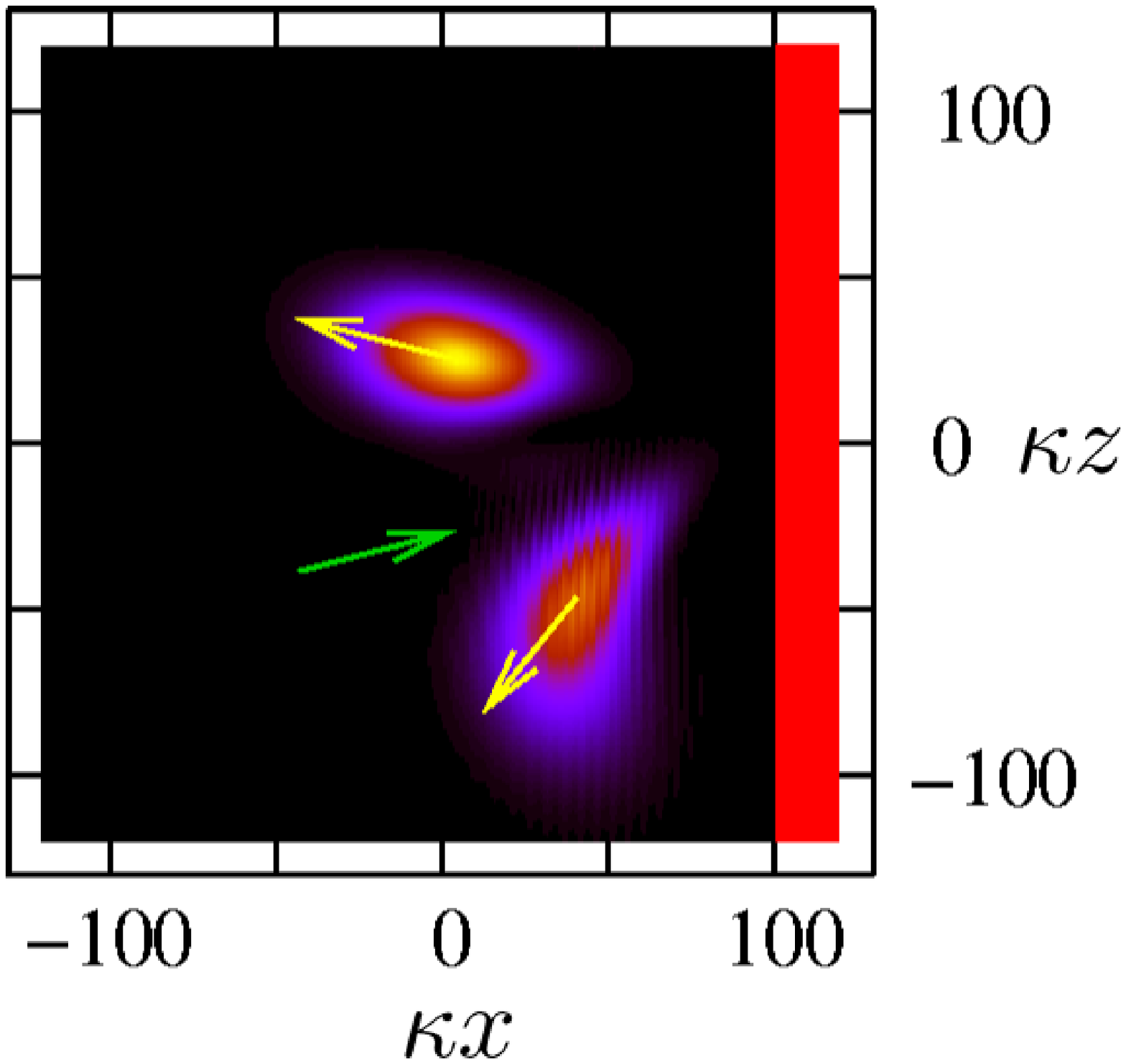}
\includegraphics[width=0.23\textwidth]{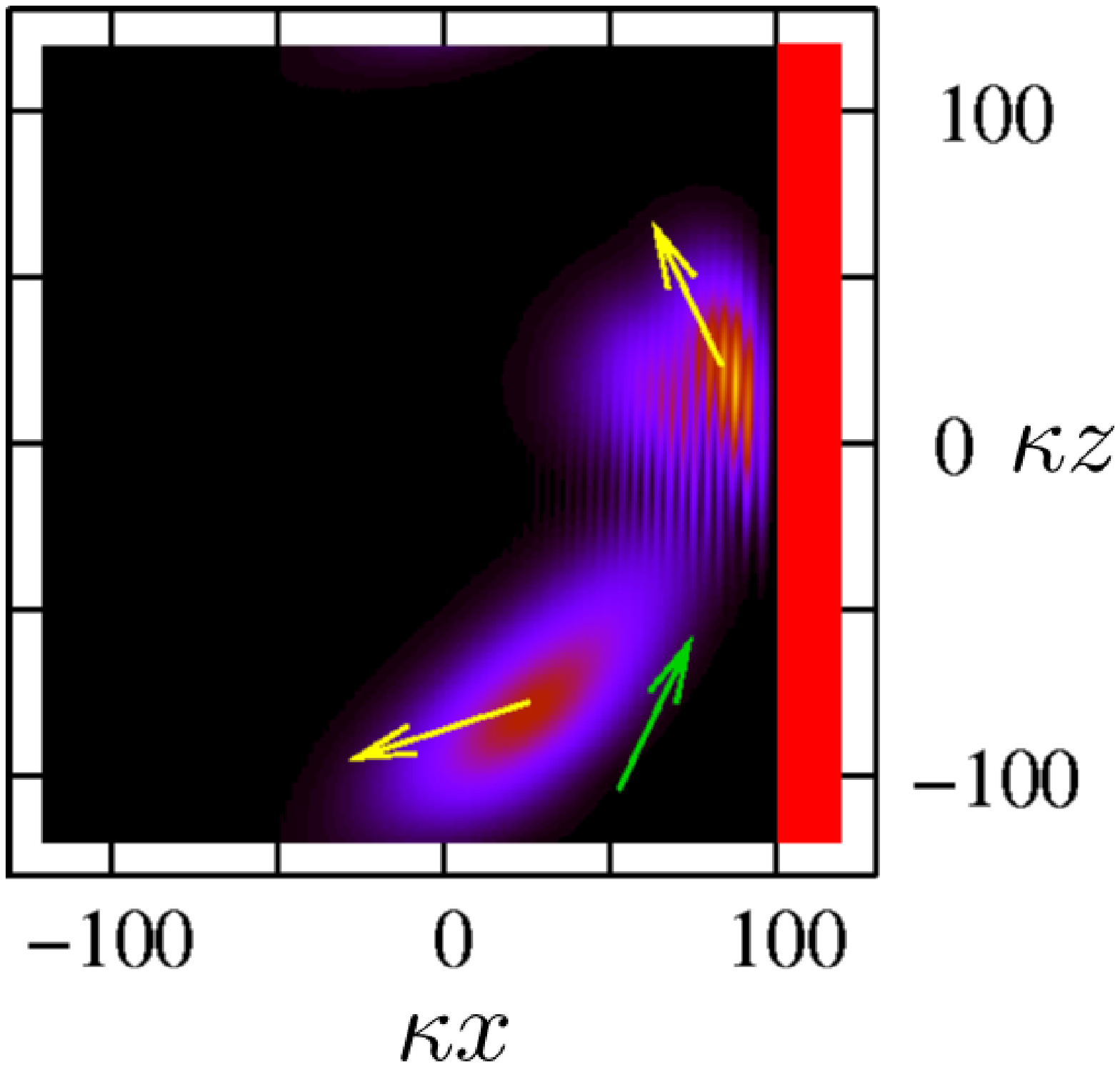}
\caption{(Color online) Reflection of an atomic wave-packet with a negative
chirality for $\alpha=15^{\circ}$, $k=1.5 \kappa$  (left) and
$\alpha=65^{\circ}$, $k=0.5 \kappa$  (right). The incident wave-packet is taken
to be Gaussian with momentum width $\Delta k=0.1 \kappa$. An additional arrow
indicates the incident direction.}
\label{fig:wavepacket}
\end{figure}

Our plane-wave analysis may be easily extended to the case of wavepacket
reflection. Similar results may be found if the momentum width of the
wavepacket $\Delta k$ is sufficiently small with respect to $\kappa$.
Fig.~\ref{fig:wavepacket} displays the double and negative reflection of atomic
wavepackets from an atomic mirror,  for an incident wavepacket
$\Psi\left(\mathbf{r}\right)=g_{\mathbf{\bar k}}^{-}e^{i\mathbf{ \bar
k}\cdot\mathbf{r}} f(r)$, with $f(r)$ a Gaussian, and $\mathbf{\bar k}$ the
central wavenumber.  The propagation direction and population of each of the
reflected wave-packets are in good agreement with the analytical plane-wave
results (\ref{eq:alpha-2})-(\ref{eq:P-1-P-2}).  Similar results are also found
for more realistic Gaussian or evanescent atomic mirrors 
\cite{OPT,MAGN}, as long as the
potential barrier is sufficiently high compared to the incident kinetic energy.
Lower barriers would lead to partial reflection, transmission and tunneling,
whose physics will be the subject of further investigations. 

Summarizing, the reflection of atoms under a non-Abelian gauge potential
presents unusual features. In particular, one can have a double reflection
comprising a specular and a non-specular one.  Remarkably, the latter wave
shows negative reflection due to the special properties of the dispersion law,
and becomes evanescent for sufficiently large incident angles. Atom mirrors are
a key tool in atom optics. Hence the anomalous reflection properties may be of
crucial importance for the design of non Abelian atom optics elements, e.g.
atom interferometers which exploit the non-Abelian Aharanov-Bohm effect.

\begin{acknowledgments}
This work was supported by the Royal Society of Edinburgh, the UK Engineering
and Physical Sciences Research Council, the Deutsche Forschungsgemeinschaft
(SFB407, SPP1116), and the European Graduate College on Interference and
Quantum Applications.
\end{acknowledgments}

\end{document}